# Sentiment Analysis for Arabic in Social Media Network: A Systematic Mapping Study

Mohamed Elhag M. Abo[1]. Ram Gopal Raj[1*]. Atika Qazi[2], Abubakar Zakari[3]
1. Department of Artificial Intelligence University of Malaya Kuala Lumpur, Malaysia.
2. Centre for Lifelong Learning, University of Brunei Darussalam, Gadong, Brunei Darussalam.
3. Department of Computer Science, Kano University of Science and Technology Wudil, P.M.B 3244, Kano, Nigeria.

Corresponding author: Ram Gopal Raj (e-mail: ramdr@um.edu.my). Author email 2: aboo72me@gmail.com.

## Abstract

With the expansion in tenders on the Internet and social media, Arabic Sentiment Analysis (ASA) has assumed a significant position in the field of text mining study and has since remained used to explore the sentiments of users about services, various products or topics conversed over the Internet. This mapping paper designs to comprehensively investigate the papers demographics, fertility, and directions of the ASA research domain. Furthermore, plans to analyze current ASA techniques and find movements in the research. This paper describes a systematic mapping study (SMS) of 51 primary selected studies (PSS) is handled with the approval of an evidence-based systematic method to ensure handling of all related papers. The analyzed results showed the increase of both the ASA research area and numbers of publications per year since 2015. Three main research facets were found, i.e. validation, solution, and evaluation research, with solution research becoming more treatment than another research type. Therefore numerous contribution facets were singled out. In totality, the general demographics of the ASA research field were highlighted and discussed.

## 1. Introduction

With the persistent increase in digital text resources in social media networks, Web, and Organizations, Sentiment Analysis (SA) techniques have become necessary for supporting knowledge discovery. Text sources and SA applications are different. Even though there is not a consensual definition set among the many research communities [1], SA can be determined as a set of processes used to determine unstructured data and know emotions that were unknown beforehand [2].

The Arabic language is the common rich active part of the group of Semitic languages regarding speakers. It is roughly related to Amharic and Aramaic languages. Arabic is delivered by neighbouring 400 million populations living in North Africa, the Middle East, and the Horn of Africa [3]. The Arabic language appears in three classes, which are Dialect Arabic (DA), Modern Standard Arabic (MSA), and Classical Arabic (CA) [4]. MSA is used in formal situations, such as news reports, classrooms, and marketing forums. CA is found in religious texts, which is primarily used in prayers and for reading the holy Qur'an.

Besides, DA is used in everyday life in comments or discussions posted on social media networks. Considering the growing significance of the study population in Arabic sentiment analysis (ASA), a raise of pertinent techniques had to stand out over the last decades, and many types of research have been reported in this area [2, 5-7]. Furthermore, SMS in this research field is nonexistent. This SMS moves to fulfil this gap.

In this work, we conducted an SMS to investigate the existing literature on SA research domain for social media networks such as blogs, Twitter, Facebook, and the Arabic language. To ensuring transparently and cover of all appropriate research, and evidence-based systematic mapping protocol is selected, which ensures coverage of state-of-the-art research by following a systematic and a fair selection and evaluation method [8]. This study is started with the development of systematic mapping rules comprising of a search strategy, inclusion/exclusion patterns, study selection operation, data extraction, and data synthesis plans. Hence, we present the SMS results by integrating evidence into patterns that can be used to understand the current state-of-the-art of research in social media SA for the Arabic language. Generally, this SMS investigates the overall research output, demographics, and trends shaping the landscape of this research area. Findings from the analysis are presented. Hence, research gaps and trends in the existing body of knowledge are also highlighted. Our findings will aid both new and veteran researchers in the field of study to understand ASA for social media.

Our contributions in this paper are as follows: Firstly, to provide transparency and composition of all related studies, an evidence-based systematic mapping methodology was raised, which guarantees the coverage of state-of-the-art review by accepting a systematic, fair selection, and evaluation method.

Secondly, we present a detailed analysis and overview of ASA studies according to seven aspects (technique, machine learning approaches, level of sentiment analysis, Research facets, contribution facets, applications platforms, and demographic characteristics). Thirdly, this is the most comprehensive and up-to-date systematic mapping based on the papers it covers on systematic Arabic study covered modern standard Arabic and Arabic dialect. Lastly, we provide a starting point for researchers and practitioners who are working on ASA.

This paper is organized into five sections. Section 2 presents related works. Section 3 highlights the research method. In Section 4 the results and discussion are presented in detail. Lastly, the study is concluded in Section 5.

## 2. Related Work

This section highlights the existing survey and review papers available in the ASA research area with the importance and necessity for our contribution to the research area. In a paper [4], the authors present an inclusive overview of the studies done so far in ASA. Moreover, the author's collections published articles based on SA related problems. Furthermore, they addressed and attempted to identify the gaps in the review, laying the basis for future researches in this field. Hence, limitations were highlighted with existing research problems in the field of study. In another study [9], the authors examined the SA conducted in Arabic Internet information sources. The authors include studies based on corpus method, which comprises Arabic Internet information sources. The study deals with both MSA and DA. However, the work was limited to a lexicon-based approach.

In a study by [10], the authors conduct a review of the Arabic language. A deep exploration of the existing works was concentrated on dealing with three approaches supervised, unsupervised, and hybrid. The findings are interesting but divergent. Hence, some approaches and techniques were not covered in this study . Furthermore, in a study by [11], the authors surveyed various studies related to Arabic text mining with more emphasis on the Holy Quran, SA, and web documents. However, this research covers a few problems and methodologies. In a study by [12], a comprehensive review of ASA was conducted. The authors give the advantages and disadvantages of various methods used and highlight their challenges. The authors also outline the relevant gaps in the research and give recommendations for future research. In a study by [13], a survey of some of the most important current lexicon, machine learning, and hybrid sentiment classification methods for the Arabic language was conducted. However, looking at all the highlighted survey studies, there is no systematic mapping research in ASA research area. The review choice process of existing survey papers is inconsistent, with no rigorous and repeatable evidence-based paper selection process. Also, there are no survey studies in ASA research domain that categorize studies in terms of their research/contribution facets, publishing trends/forums, most-used datasets and utilised evaluation metrics, citation impacts, authorship information, and geographical distribution of selected studies. Then, this research aims to fill these research gaps.

## 3. Research method

Mapping studies are conducted to give a general overview of a specific field of study by classifying and quantifying research contributions with regards to the classifications made [14]. In this paper, we adopt the systematic mapping method represented by  Gough et al. [15]. This mapping process is adopted for our research purpose because our objective is to explore the existing studies related to ASA. The results of the mapping study would help us find and map the ASA research area and further identify research gaps. The process for the SMS is shown in Figure 1, which content of five process steps and their output.

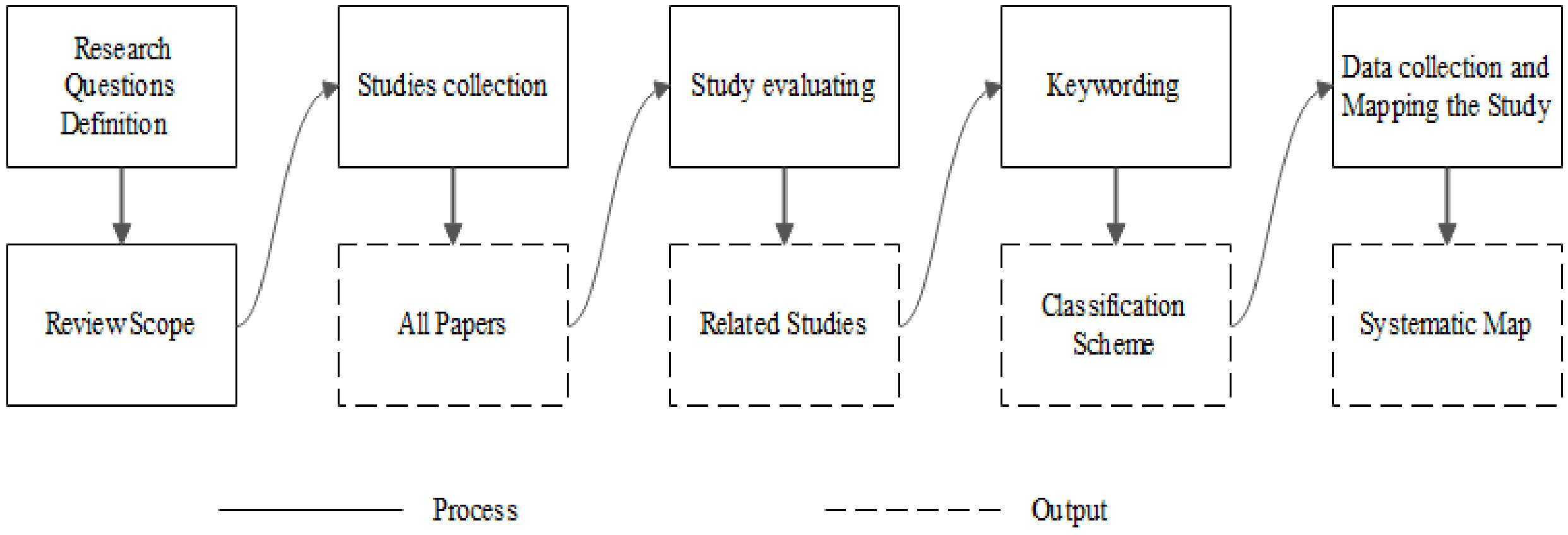

Fig 1. The Systematic Mapping Process

## A. Research Questions

In this section, the research questions (RQs) for this study are outlined. The main RQ of this SMS is: "What is the state-of-the-art in ASA studies?" This RQ is broken into seven RQs, as pointed up in Table 1, along with the motivation of each various RQ.

Table 1. Research Questions

| | Research Questions | Motivations |
|---|---|---|
| RQ 1 | What are the sentiment analysis techniques used in Arabic language studies? | To identify the commonly Sentiment Analysis techniques used in Arabic analysis. |
| RQ 2 | What are the sentiment analysis approaches used in Arabic language studies? | To identify the approaches commonly used in ASA. |
| RQ 3 | What are the levels of sentiment analysis that are used in the Arabic domain? | To identify the sentiment analysis levels commonly used in Arabic text classifications. |
| RQ 4 | What research approaches (Research facets) and contribution facets have been focused on sentiment analysis studies and what do they provide? | To distinguish the research facets (Solution, evaluation, philosophical, validation, etc.) and contribution facets (technique/method, tool, framework, evaluation/comparison, and so forth.). |
| RQ 5 | Which datasets applications platforms are widely used in ASA research community? | To find an aggregated list of relevant datasets platforms for ASA research. |
| RQ 6 | What is the most used evaluation metrics? | To highlights the most utilized evaluation metrics based on the SA techniques used in the Arabic research area. |
| RQ 7 | What are the demographic characteristics of the relevant studies? | To cover the temporal and geographical distribution of primary selected studies, relevant publication venues, highly cited studies, influential authors, and active research groups in the field of research. |

## B. Data sources

SA is widely praised in the academic review and has newly acquired considerable popularity [4]. The following Four digital databases listed in Table 2 were used to seek for target articles.

Table 2. Show database website

| ID | Name | Description | Links |
|---|---|---|---|
| DW 1 | Web of Science | Service, an indexing database that covers different academic disciplines. | http://apps.webofknowledge.com |
| DW 2 | ScienceDirect | Science and technical journal articles. | http://sciencedirect.com/ |
| DW 3 | IEEE Xplore | Library of technical literature in engineering and technology. | http://ieeexplore.ieee.org/ |
| DW 4 | ACM | Association for Computing Machinery | http://dl.acm.org/ |

### C. Search Terms

The identification of search terms is vital in order to find essential studies. A search term is basically the combination of characters entered by a researcher into a search engine to find desired results. Therefore, the outcome of the search engine impacts the results provided by it. Hence, care has to be taken when selecting these keywords and characters in our search terms. The keywords used for our search are ("sentiment analysis") AND (Arabic) AND ("Social media"). Moreover, in this paper, we only covered Arabic language SA studies and the general category of SA domains.

### D. Search result

As shown in Figure 2, the initial query search resulted in 277 articles published in 5 years period (2015-2019). The Figure shows that 69 articles are from Web of Science, 147 from Science Direct, 51 from IEEE Xplore, and ten articles from ACM. Moreover, based on our five years article selection criteria, a total of 236 articles were obtained. We further get 204 articles after we filter and screened out duplicates across all the four libraries.

Furthermore, by filtering out paper types (article or conference paper) across all four libraries, we removed reviews papers, surveys papers, and book sections/chapters. With that, we get 146 articles, as shown in Figure 2. We further filter the 146 articles by their publication titles and abstracts, which we excluded 49 articles across all libraries in this stage. Lastly, 51 final selected articles for all the selected databases were obtained as a final set of studies targeted for analysis.

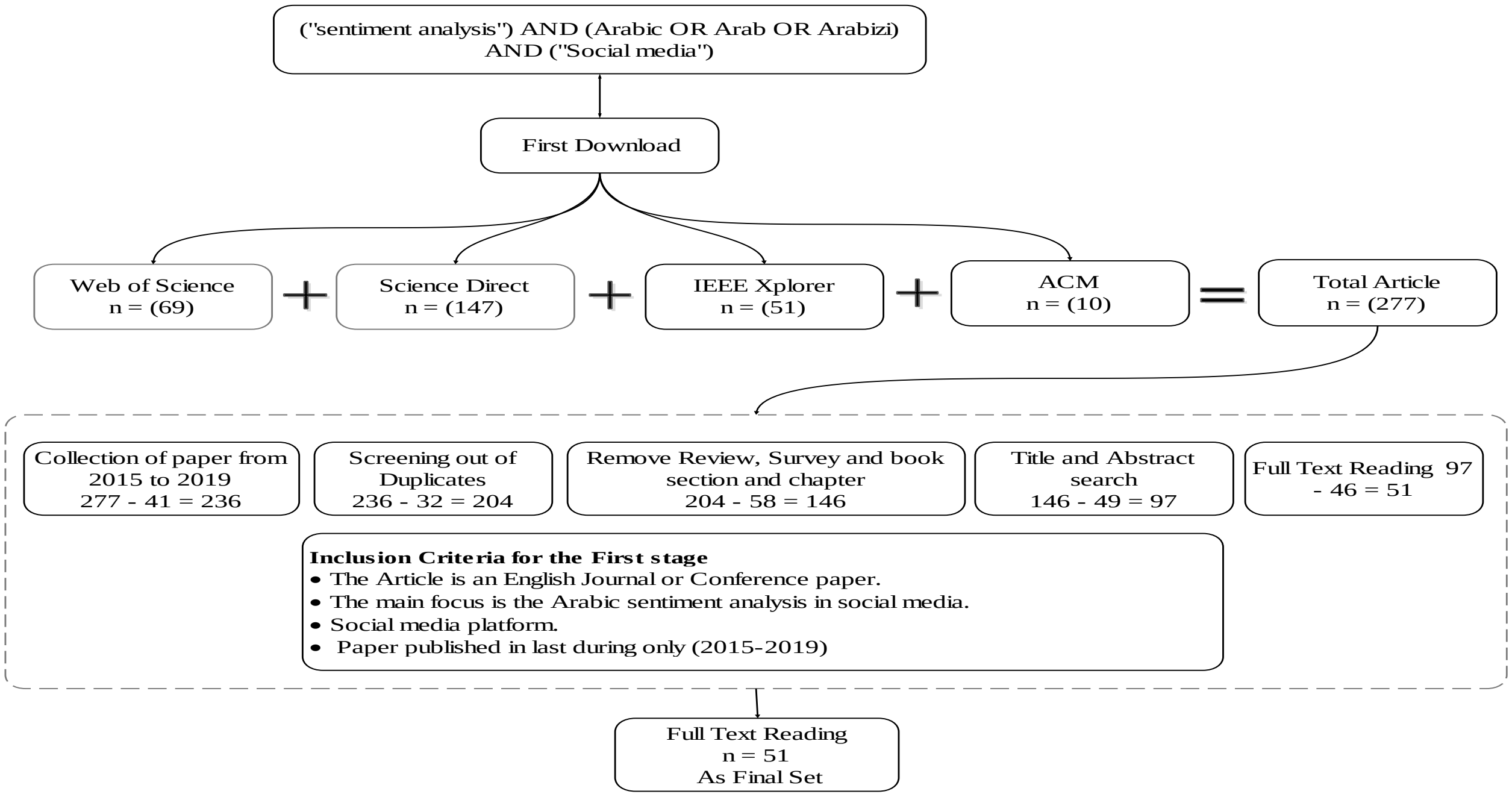

Fig. 2. The primary paper selection process

## E. The Exclusion and Inclusion Criteria

exclusion and Inclusion criteria were used to select potentially relevant research papers from the data sources to answer the relevant research questions in a given systematic mapping study. These conditions were applied to all papers retrieved in the different phases of the study selection procedure (see Fig 2.). After obtaining all the primary selected studies, a well-designed exclusion and inclusion criteria were applied to these studies to eliminate articles that are not in-line with the objectives of this mapping study. The exclusion and inclusion criteria employed in this SMS are in Table 3.

Table 3. Inclusion and exclusion criteria

| | |
|---|---|
| Inclusion criteria | |
| 1 | ASA that targeted social media. |
| 2 | The article is an English Journal or Conference paper. |
| 3 | The studies published within five years the period of 2015-2019. |
| Exclusion criteria | |
| 1 | Review and survey papers. |
| 2 | Studies that are not based on opinion mining or sentiment classification. |
| 3 | Non-text form (audio or video classification). |
| 4 | Books, sections and book chapter. |
| 5 | Papers that do not target social media text. |

## F. Classification Scheme

We have presented a characterization plan for ASA schemes, as showed in Figure 3. This characterization plan incorporates both special highlights and perceptions from the existing benchmark examines. It comprises of four essential dimensions: (1) Commonly used Arabic SA techniques, (2) Research facets, (3) Contribution facets, (4) Commonly used evaluation metrics, (5) Level of sentiment analysis, and (6) Commonly used datasets.

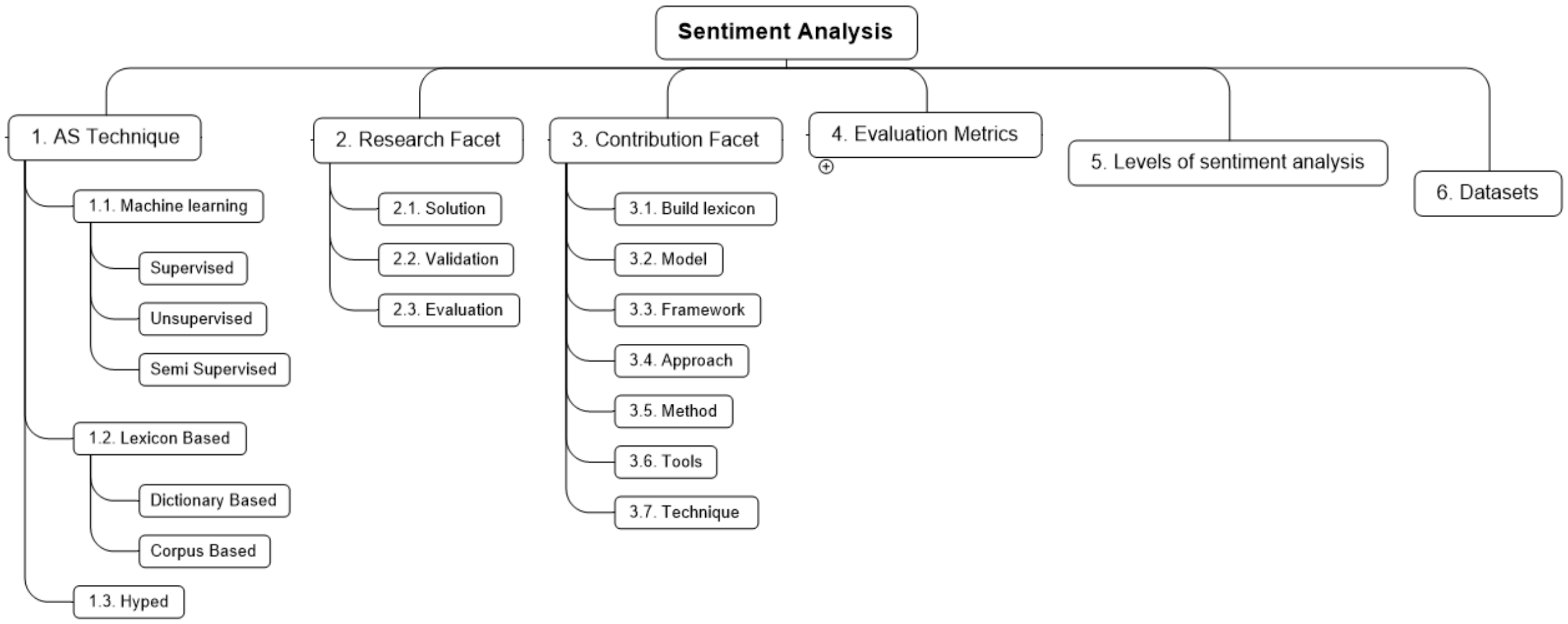

Fig 3. Classification for Systematic Mapping Study

In the context of SA techniques for Arabic text, we have created a list of three main categories, which are hybrid, machine learning-based, and lexicon-based. Moreover, there are subcategories under machine learning. These subcategories are a supervised approach, unsupervised approach, semi-supervised approach, and hybrid approach.

For *Supervised approach*, all data are labelled, and the algorithms learn to figure out the output from the raw data [16]. Moreover, for the *unsupervised approach*, all data are unlabeled, and the algorithms learn to inherent structure from the input data [17]. Furthermore, some data are labelled, but most are unlabeled; hence, the mixture of supervised and unsupervised approaches can be used, which is called *semi-supervised* [18]. However, *the hybrid approach* is when machine learning and lexicon-based approach are combined in ASA [16].

Moreover, *Lexicon-based* is classified into two approaches, which are Dictionary-based and corpus-based. The dictionary-based sentiment analysis approach uses word similar based on the lexicon [18]. Moreover, *corpus-based* is a data-driven approach where it will access not only sentiment labels but to a context which can be used to advantage in a machine learning algorithm [9, 19].

The second classification is to explore and classify primary selected studies correlate with the current research methods. Moreover, there are three subcategories under research facets, which are evaluation research, validation research, and solution research [9]. *Evaluation research* seeks to implement techniques in practice; an example of evaluation research is [7, 20-25]. Evaluation of machine learning performance for ASA was conducted in [26]. Such type of research facet illustrates the applicability of the technique in usage along with its advantages and disadvantages. *Validation research* is the investigation of novel techniques, method, model, etc. in SA that has not yet been implemented in practice, it can be experiments conducted in a lab or controlled environment [27]. Lastly, for solution research, this research facet suggests a solution to a problem that can be novel or a significant continuation of an existing technique in ASA [28].

Thirdly, contribution facets are technical benefactions identified from primary selected studies on ASA. There are seven subcategories under contribution facets, which are Build lexicon, Sentiment analysis Model, framework, Approach, Method, Technique, and tools.

*Build lexicon* is built a database that contains the famous positive and negative words that can help to implement sentiment analysis in Arabic text [16]. *Sentiment analysis Model* is a way of customizing the sentiment analysis carry out in texts. There are different scenarios of model-based on the specific language used or because of the analysis type are trying to achieve. *The framework* is a qualitative method that is aptly suited for applied policy research for SA or opinion mining [29]. For *Approach/Method/Technique*, an approach is a set of correlative assumptions about the nature of language sentiment analysis. However, a method is a plan for presenting the sentiment analysis steps to be conducted and should be based upon a selected approach. Furthermore, a technique is a particular concrete scheme or method designed to accomplish SA [4, 30-32]. Moreover, *sentiment analysis tools* are software, code, or computer program that are used for SA purposes [32].

Fourthly, for *evaluation metrics*, there are various performance metrics used to assess different machine learning-based algorithms. *Accuracy, F-measure, Area under Curve (AUC),* and *F1- Scour* is used for classification problem in machine learning for ASA. Moreover, another example of metrics used for evaluating machine learning algorithms is *recall* and *precision*. These measurements can be utilized for arranging algorithms which are principally utilized via search engines. *The receiver operating characteristic (ROC)* which is described as a plot of test sensitivity as they coordinate versus its 1-specificity or false positive rate (FPR) as the x organize, is a powerful method for assessing the performance of analytic tests. *Bilingual evaluation understudy (BLEU*) is a process for assessing the quality of text which has been Auto-translated from one language to another. Quality is the correspondence among a machine's output and that of a human. *Geometric Mean (GM)* is a type of average, frequently used for growth rates like population growth or interest rates. *Out-Of-Vocabulary (OOV)* is a speech recognition assessment metric; the accessible vocabulary should initially be defined. *OOV* refers to the jargon contained in the input audio signal, which isn't a piece of the accessible vocabulary lexicon. Therefore, it will always be miss-recognized using automatic speech recognition. *Cross-Validation* is also utilized for avoiding the issue of over-fitting which may emerge while planning a supervised classification model like Artificial neural systems (ANN) or Support vector machine (SVM). It is a technique that can give the correct accuracy of a model. *The gold standard (gold-std)* is an accepted standard that people can look to as an accurate and reliable reference[33]. *Matthews correlation coefficient (MCC)* is also used in machine learning as a measure of the quality of binary (two-class) classifications[34].

### G. Data Extraction and Systematic Map

All research papers selected after the screening phase were analyzed by the review groups. The full texts of all qualifying articles were analyzed by at least two academics. Moreover, related information was extracted to a predefined data extraction form. Figure 4 explains the data extraction form used to extract data items. Related fragments of the studies where useful information such as the application of datasets and evaluation metrics were highlighted. This helped us in quickly detecting and validating data extraction results and resolving disagreements. Extracted data items are then synthesized, and the results were described accordingly.

Two specialists RAM GOPAL RAJ and ATIKA QAZI considered and planned the designed and directed the research as indicated by the defined mapping protocol; extracted and analyzed the information as two review group. MOHAMED ABO and A ZAKARI coached and guided the experimentation and overall research process. The researchers directed independent information extraction and categorizations. Hence, these results were collected in a revision meeting including all researchers. During the meetings, the results of data extraction and categorization process were deemed not to have any conflict or contradiction.

**Data Collection Form**

| | |
|---|---|
| Reference No | |
| Article Title | |
| Author | |
| Year of publication | |
| Paper type | ☐ Journal ☐ Conference |
| SA Technique | ☐ ML ☐ lexicon Based ☐ Hyper |
| Domain (service/finance/book review) | |
| language | ☐ MSA ☐ AD ☐ Unformal |
| Dataset application (Twitter/Facebook etc.) | |
| Sentiment analysis level | ☐ Doc ☐ Sentence ☐ word/aspect |
| Methods/Algorithms | |
| Features | |
| Approaches | ☐ Supervised ☐ Unsupervised ☐ Semi supervised |
| Dataset | |
| Analysis software (R – python etc.) | |
| Metrics | ☐ Accuracy ☐ Recall ☐ Precision ☐ F-measure<br>☐ Other ................................ |
| Contribution facet | 1- Development<br>2- Enhance<br>3- Propose model/Tools<br>4- Review result<br>5- Evaluation<br>6- Build lexicon<br>7- comparison |
| Inclusion/ Exclusion | ☐ Inclusion ☐ Exclusion |
| Note | |

Fig 4 Data extraction form

# 4. Result and Discussion

This section highlights the key findings of our systematic study. The RQs answered by analyzing the results collected from our primary selected papers.

## A. Techniques used in Arabic language studies (RQ1)

The results of the analysis that have been done on selected primary studies, we identified three prominent ASA methods which are depicted in Figure 5. We observed that 76% of the papers adopt machine learning-based technique, 14% are based on lexicon-based techniques, and 10% is based on hybrid techniques. Moreover, lexicon-based methods that are part of the SA approaches listed in Figure 5. Moreover, the techniques have two types of method; one is evaluating existing ASA lexicon, which contributes to three papers of our selected studies [5], [35], [23].

Furthermore, another method is building lexicon for Arabic such as modern standard Arabic or Arabic dialect which has four Papers [36-39]. Hybrid ASA techniques as categorized in Figure 5 are techniques that combine more than one SA techniques to improve the identification of sentiment in Arabic text. From our collected studies, we have seen an increase in the utilization of hybrid SA techniques in the last few years, which contributes to 10% of the techniques being utilized from our primary selected studies.

Moreover, as we can see in Figure 5, the primary selected studies are classified based on publication years. In 2015, there are six (12%) papers with three studies using machine learning-based technique and two papers using the lexicon-based technique. However, only one paper utilized the hybrid approach. Moreover, in 2016, the number of studies increased with six papers (14%). Most of these papers are using machine learning algorithms for ASA. However, only one study uses the lexicon-based technique. In 2017 and 2018, ASA got attention, which is increased about two times as we can see in Figure 5. Moreover, the percentage is 31% and 27% respectively, which is about 30 paper in our selected area. 2019 also showed 16% with eight papers from selected studies, which is seven papers using machine learning and one study use lexicon-based. Moreover, 2019 is a promising year for ASA studies, because, in the middle of the year, there are more than 50% of publications of ASA in comparison with the previous year (2018).

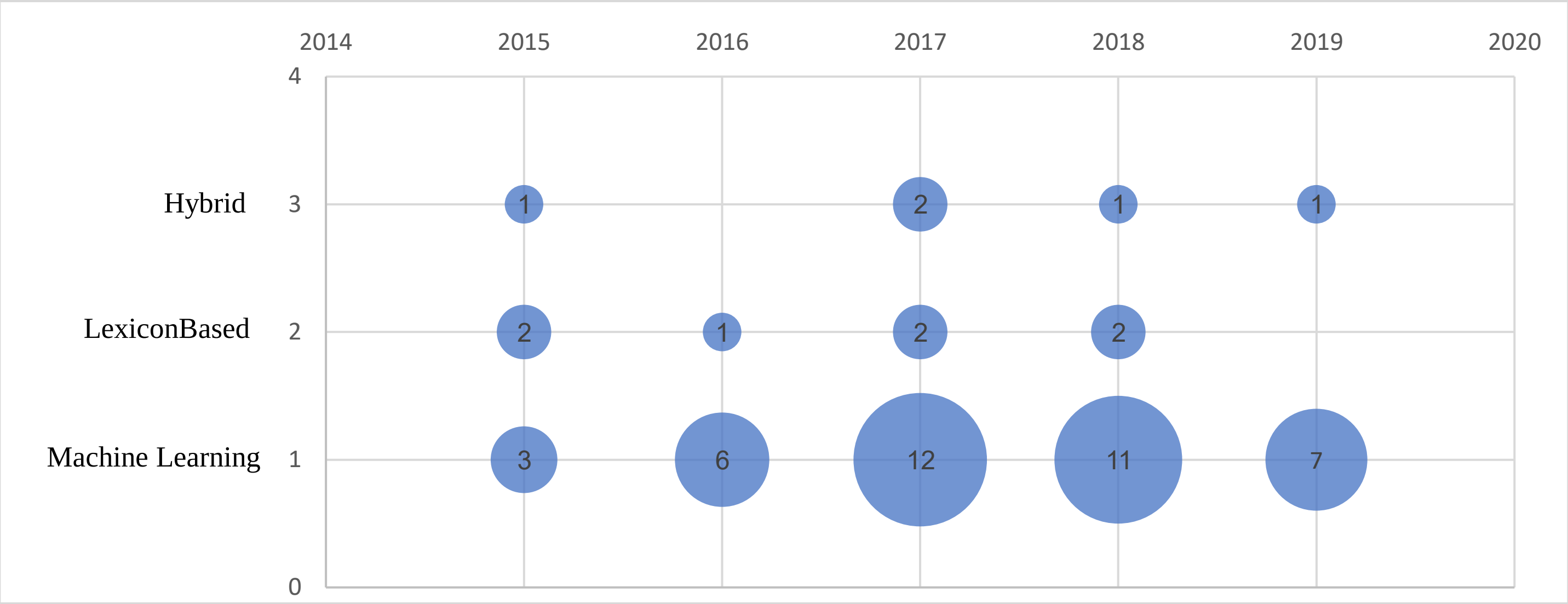

Fig 5. ASA Techniques proportion by years

## B. Approaches used in Arabic SA studies (RQ2)

Supervised learning approach with regards to artificial intelligence and machine learning is a sort of methodology where both input and desired output are given. Input data and output information are annotated for classification to give a learning basis to future data processing [2, 40]. Moreover, the semi-supervised approach is a class of ML-based technique that also allows the use of unlabeled data for SA. Training data usually set a small volume of labelled data with a large amount of unlabeled data for the testing set. The semi-supervised approach used in ASA as categorized in Figure 6 is an approach that combines learning approach with unsupervised ASA to improve the identification of the sentiment in Arabic text. However, in view of our examination of the chose primary studies, we identified three prominent machine learning approaches which are depicted in Table 4. These approaches are supervised, unsupervised and semi-supervised approach. We observed that 70% of the selected papers use a supervised approach which amounts to 36 papers of the total selected papers; the example is proffered in [20, 26]. Furthermore, 18% of the selected studies used the unsupervised approach which is nine studies, and the example showed in [17, 40], while the semi-supervised approach has 12% which is only six papers the example is provided in [41, 42]. From our chose papers, we have seen the expanding utilization of semi-supervised sentiment analysis approach over the most recent couple of years, which contributes over 10% of the all-out approaches being used in the research field.

Table 4. SA approaches in Arabic studies

| Rank | Research approaches | Primary studies |
|---|---|---|
| 1 | Supervised Approach for Arabic SA | [6, 8, 18, 20-26, 35, 36, 43-64] |
| 2 | Unsupervised Approach for Arabic SA | [7, 17, 37, 38, 40, 65-68] |
| 3 | Semi-supervised Approach for Arabic SA | [5, 41, 42, 69-71] |

### C. Sentiment analysis levels used in Arabic domain (RQ3)

The selected papers were classified based on the level and approach of SA used, which is depicted in Figure 6. In this study, we identified three main SA level which are a sentence, document, and word/aspect level. Word level in supervised approach got the highest number of studies in our selected papers with 29 (56%) papers. An example of the word level under a supervised approach is given in [36, 49]. Moreover, the unsupervised approach is displayed in 6 papers, which amount to 12% of the selected studies. An example of the word level under unsupervised approach is given in [5, 71]. Also, the semi-supervised approach has four papers for word-level, which also amount to 8%. Sentence level in ASA comes after word-level SA in a supervised approach with 4 (8%) studies. Unsupervised and semi-supervised showed 3 (6%) and 2 (4%) papers for each in sentence-level of ASA, an example is given in [7, 17, 21, 66, 70]. Finally, the document level showed only four studies under a supervised approach, which is 8%. An example of the document level under a supervised approach is given in [20, 53, 57, 59].

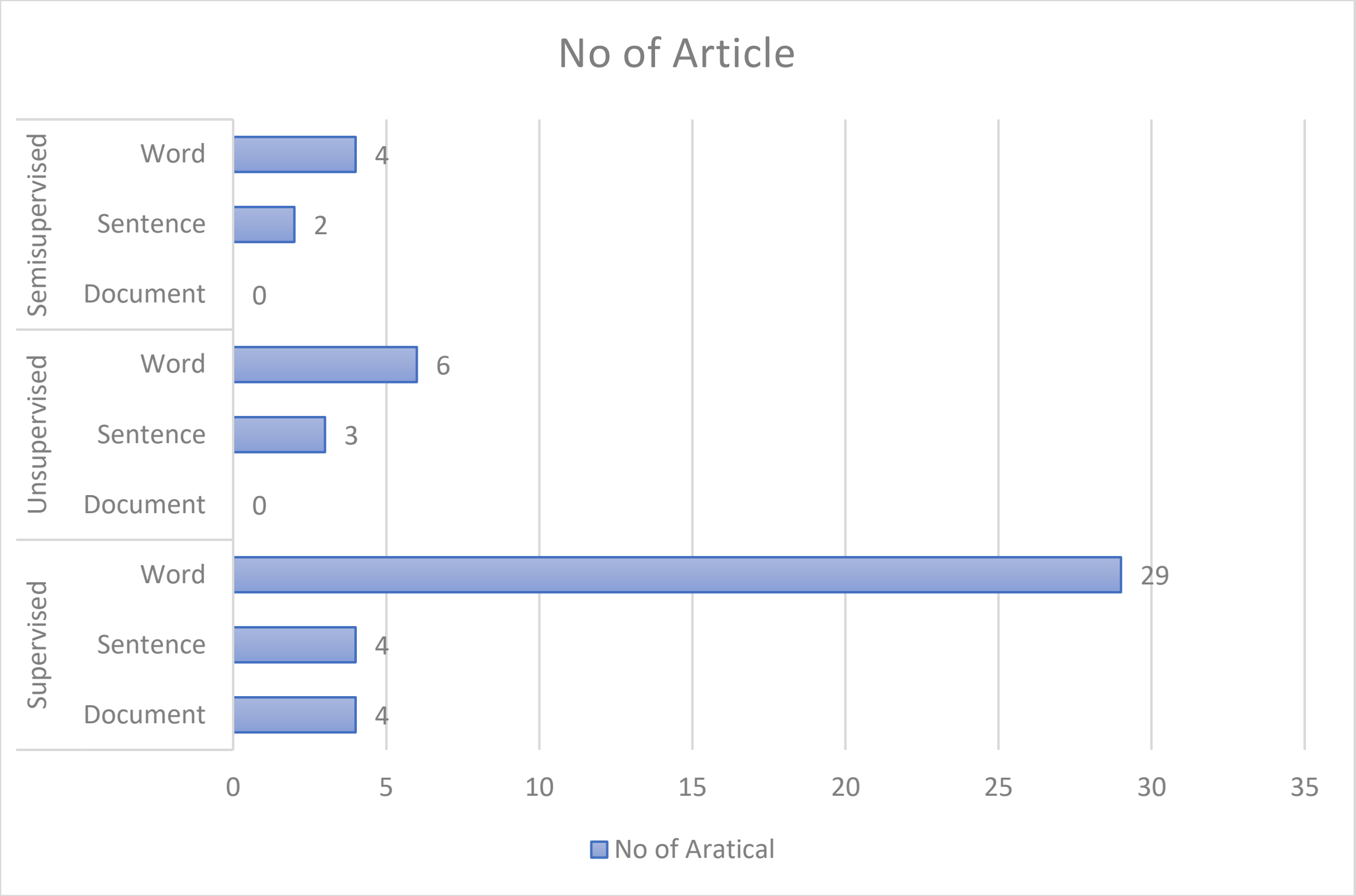

Fig 6. SA levels with ML technique

### D. Research facets and contribution facets (RQ4)

The primarily chosen research are additionally classified based on research approaches and contribution facets. In addition, contribution facets are grouped into Build Lexicon, model, Framework, approach, tool, technique, and method (See Table 5). In this mapping study, the *models'* contribution is 21%. On the other hand, contribution facet in terms of *authors'* is 4%, which makes up the least studies with only two studies [40, 63]. The *framework* refers to a comprehensive structure covering a group of steps by aiming to extract sentiment analysis from Arabic text. An example of the framework is given in [24, 25, 38, 59]. These studies provide a framework that detects ASA in social media network. In addition, comparing approaches and method, the approaches present 12%, which contribute 2% more than the method used in ASA. Moreover, six papers for method, an example of the method is given in [26, 57]. Moreover, five studies used SA approaches, an example of the Approaches is given in [64].

A web-based tool is developed and evaluated on SA various types of Arabic language such as modern standard Arabic or Arabic dialect[55]. Moreover, Doc2Vec tool is used for Arabic opinion mining [70], and the tools are presented in 11 studies is 22 %. In comparison, a *technique with tools* focuses on a goal of solving a purpose such as Arabic news effect on readers or a well-defined research question. An example of the tools/technique is presented in [63], which offered a novel BRST method for ASA. Table 5 shows the studies done in contribution to facet.

This mapping study looks to investigate changed research approaches or research facets with regards to ASA explore. Research facets were grouped into evaluation, solution and validation research. Solution research proposes an answer for an issue that can either be novel or a noteworthy augmentation of a current technique. A sample of solution research is found in [72] that recommended a method for ASA in social media platforms such as Twitter. Validation research offers novel techniques that have not yet done performed and fully assessed. An example of validation is found in [5, 42]. Finally, evaluation research tries to execute techniques and an assessment of the techniques is accessible. Such sort of research facet demonstrates the pertinence of the proposed technique alongside its favourable advantages and disadvantages. A case of evaluation research is found in [26, 45, 53]. Analysis of primary selected studies in terms of research facets reveals that most of the studies were conducted using evaluation research (49%). On the other hand, the quantity of studies performing Solution research approach is moderately little (47%). The least is validation research which contributed (4%). This means the vast majority of the current research on ASA is confined to evaluate recommendations and tests directed in a controlled environment. This suggests there is a requirement for more validation research that can assess how powerful these new ASA solutions really are. A map of research types in table 6 illustrates the distribution of research facets of ASA techniques.

Table 5. Show the Contribution facet studies of ASA

| Contribution facets | Primary studies |
| --- | --- |
| Build Lexicon | [5, 23, 35-37, 39, 43, 44, 50, 54, 67, 68] |
| Model | [7, 17, 18, 21, 22, 41, 42, 51, 60, 65] |
| Framework | [24, 25, 38, 59] |
| Approach | [47, 52, 64, 66, 69] |
| Method | [26, 46, 53, 56, 57, 73] |
| Tools | [6, 8, 45, 48, 49, 55, 58, 61, 62, 70, 71] |
| Technique | [40, 63] |

Table 6. Show the research facet references

| Research facets | Primary studies |
|---|---|
| Evaluation | [5, 7, 8, 20, 22-26, 35, 40, 41, 43, 46-48, 50, 52, 56, 57, 62, 65, 67, 68, 73] |
| Solution | [6, 17, 18, 21, 36-38, 44, 45, 49, 51, 53-55, 58-61, 63, 64, 66, 69-71] |
| Validation | [5, 42] |

### E. Datasets applications platforms used in Arabic SA community (RQ5)

ASA can get some support by using outside sources of knowledge. Ontologies, semantic networks, taxonomies, and thesauruses are knowledge sources that are generally used by the SA community. When looking at the external knowledge sources used in ASA studies (Figure 7), Twitter is the most used source. The Twitter platform source is cited by 60% of the selected studies. Twitter can be used to generate or expand the recent set of SA for the subsequent three levels of classification. Moreover, 23 papers of the selected studies using Twitter data. An example is given in [39, 44, 45], which is 60%.

The second most used source is the Facebook platform, which covers a wide area of subjects and has the value of presenting the same concept of SA levels in different languages. Facebook links and types are suitable for enriching SA [45]. Facebook is displayed as a data source in six papers, which amount to 11% in our selected studies. The third most used source is YouTube, which is also found to be one of the external knowledge sources. However, it is the most cited sources in our selected studies with 6% of papers. Some of the studies that utilized it are [46, 57, 73].

Based on the selected studies, we identified more than 13 sources used in ASA. Moreover, Trip Advisor is coming number four of must use source with only two paper (4%). An example is given in [64, 69]. Finally, the other sources such as Aljazeera, AL Arabia, Yahoo!-Maktoob, AWATIF and so on are cited as ASA source which has 19% of the total of studies. It is ten paper, one paper for each source.

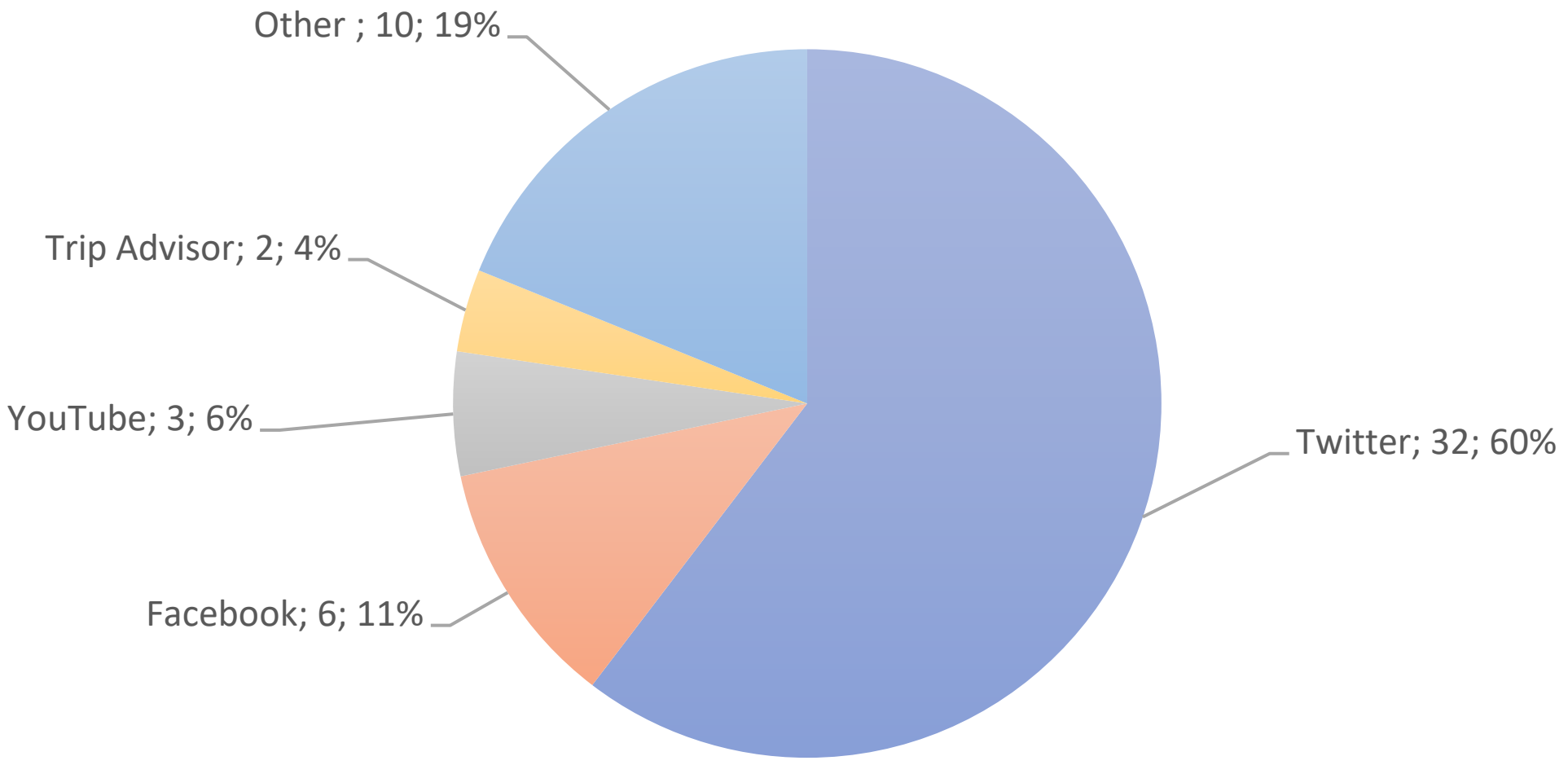

Fig 7. Applications platforms used for ASA

## F. Evaluation metrics (RQ6)

We analyze the primary studies to determine all evaluation metrics that can be utilized

To recognize the most utilized evaluation metrics in the research, we conducted a full detailed analysis of our primary selected papers to extract the used evaluation metrics. Table 7 presents all the identified evaluation metrics with the ASA techniques that utilized them. Twenty-one metrics were identified, with Accuracy being the most utilized metric, especially on machine learning-based techniques in ASA with 29% of studies using it. Followed by the recall, precision, and F-measure metrics. These metrics used more than 20 times is the ASA selected studies which show 18%, 19%, and 16%, respectively.

Table 7. Top 10 Evaluation Metrics used in ASA

| No | Metrics | No of studies | Rank | References |
|---|---|---|---|---|
| 1 | Accuracy | 35 | 1 | [7, 8, 17, 18, 21-26, 36-38, 41-50, 52-55, 57-59, 62, 65, 67, 68, 71, 73] |
| 3 | Recall | 23 | 2 | [18, 20, 21, 24, 36, 38-40, 43, 45-47, 50, 52, 54, 55, 57, 59, 60, 62, 63, 67, 69] |
| 2 | Precision | 22 | 3 | [18, 20, 21, 24, 36, 38, 39, 43, 45-47, 50, 52, 54, 55, 57, 59, 60, 62, 63, 67, 69] |
| 4 | F-measure | 19 | 4 | [21, 24, 25, 36, 38, 39, 43, 45-47, 54, 57, 59, 60, 62, 63, 67, 69, 70] |
| 5 | Receiver operating characteristic (ROC) | 3 | 5 | [21, 57, 67] |
| 6 | Bilingual evaluation understudy (BLEU) | 3 | 5 | [51, 56, 61] |
| 7 | Avg-F1 | 2 | 6 | [43, 44] |
| 8 | GM | 2 | 6 | [46, 47] |
| 9 | Area under Curve (AUC) | 2 | 6 | [46, 47] |
| 10 | F1- Score | 2 | 6 | [17, 58] |

## G. The demographic characteristics (RQ7)

To addressing demographic characteristics question five parts of the primary chose studies were analyzed, which are publication fora's that have published most relevant studies (Journals and Conference), publication trend, the geographical distribution of the primary studies, citation impacts, and authorship information in ASA field of research.

### 1. Publication Trend

From the year (2015 – 2019), 51 publications were extricated from the literature following our methodology in Section 3. Figure 8 displays the evolution of the publication in ASA literature. The research activity in ASA is progressive and active. From 2015 – 2016, the research activity was linear with few numbers of publications in conference proceedings. However, the ASA studies published journals during the year 2017 was very active, which is four times more than studies published in the conferences in the same year. Moreover, using a machine learning technique was increased significantly by improving ASA accuracy. However, in 2018, the research activity in ASA increased substantially with more than two times compared with publications in 2015.

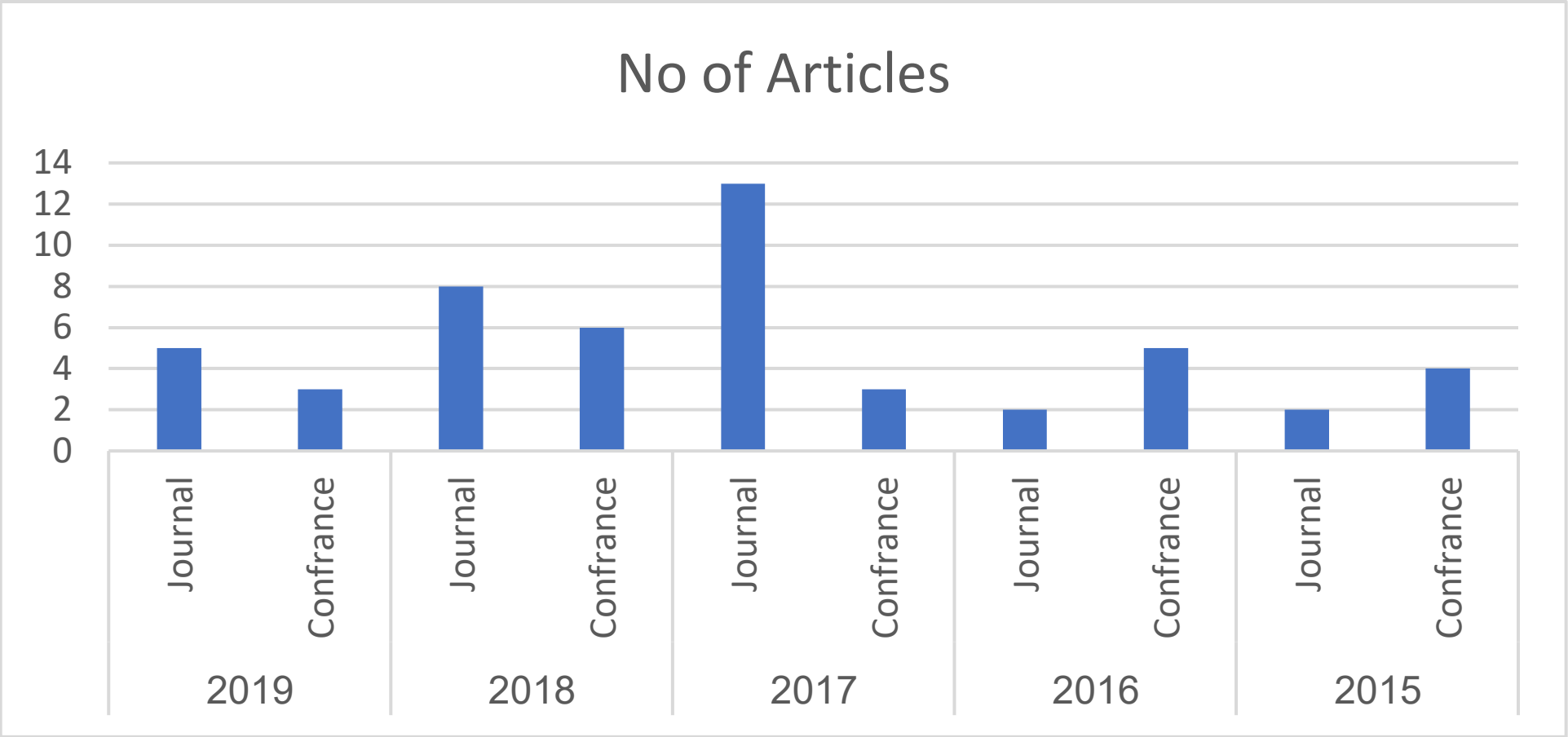

Fig 8 show publication trend

## 2. The Publication fora's

In this study, we included 30 various Journals and 21 conference proceedings. In Figure 9, most of the selected studies were journal articles (30), followed by conference papers (21).

Moreover, the book chapters, symposium proceedings, newsletter, and workshop proceedings were removed in the filtering stage in section 2. Regarding the publication venues where ASA studies were distributed, Table 8 demonstrates the main 10 most active journals. The Journal of Procedia Computer Science and International Journal of Advanced Computer Science and Applications (IJACSA) were the high contributors among every one of the Journals with 13 and six publications, respectively.

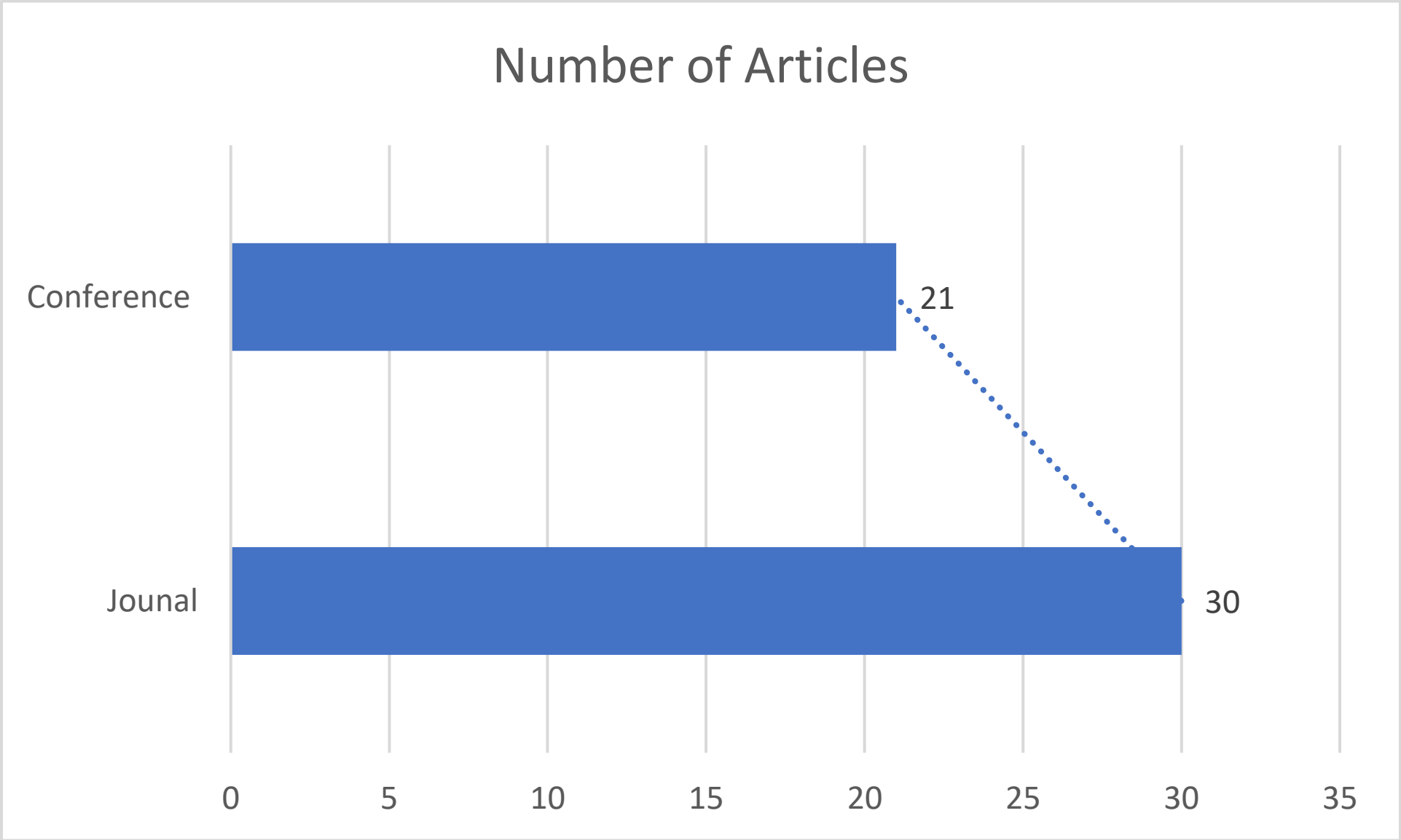

Fig 9 Publication Channel

Table 8. Top 10 most Represented Journals

| No | Name | No of paper | References |
|---|---|---|---|
| 1 | Procedia Computer Science | 13 | [5, 8, 17, 22, 23, 35, 39, 42, 48, 53, 55, 56, 64] |
| 2 | (IJACSA) International Journal of Advanced Computer Science and Applications | 6 | [18, 21, 50, 51, 66, 73] |
| 3 | Information Processing and Management | 1 | [43] |
| 4 | Expert Systems with Applications | 1 | [40] |
| 5 | ACM Transactions on Asian and Low-Resource Language Information Processing (TALLIP) | 1 | [8] |
| 6 | Egyptian Informatics Journal | 1 | [44] |
| 7 | Journal of King Saud University – Computer and Information Sciences | 1 | [36] |
| 8 | International Journal of Advanced and Applied Sciences | 1 | [52] |
| 9 | International Journal on Semantic Web and Information Systems | 1 | [24] |
| 10 | Pattern Recognition Letters | 1 | [25] |

Table 9. Top 10 most Represented Conferences

| No | Name | No of paper | References |
|---|---|---|---|
| 1 | 2016 7th International Conference on Computer Science and Information Technology (CSIT) | 2 | [60, 62] |
| 2 | 2019 International Conference on Electronics, Information, and Communication (ICEIC) | 1 | [7] |
| 3 | SMC '19 Proceedings of the New Challenges in Data Sciences: Acts of the Second Conference of the Moroccan Classification Society | 1 | [41] |
| 4 | 2019 IEEE Jordan International Joint Conference on Electrical Engineering and Information Technology (JEEIT) | 1 | [45] |
| 5 | 2018 International Conference on Computer, Control, Electrical, and Electronics Engineering (ICCCEEE) | 1 | [26] |
| 6 | 2018 International Conference on Computing Sciences and Engineering (ICCSE) | 1 | [46] |
| 7 | 2018 1st International Conference on Computer Applications & Information Security (ICCAIS) | 1 | [47] |
| 8 | 2018 21st Saudi Computer Society National Computer Conference (NCC) | 1 | [6] |
| 9 | 2018 17th IEEE International Conference on Machine Learning and Applications (ICMLA) | 1 | [65] |
| 10 | 2018 2nd International Conference on Natural Language and Speech Processing (ICNLSP) | 1 | [69] |

However, in this systematic mapping paper, we ignored a book chapter, workshop, symposium, and newsletter. We only categorize conference papers which are portrayed in Table 9. Table 9 demonstrates the top 10 most active conference for ASA. In this manner, the International Conference on Computer Science and Information Technology (CSIT) is the top contributor among the conferences with two publications.

### 3. Citation Impact

Table 10 demonstrates the number of citations of the most cited papers from our chose research papers. The citation counts of each paper obtained from google scholar citation count, which is liable to change anytime. Overall, the most cited paper is [61] with more than 50 citations. However, we have two studies with more than 20 citations each [25, 38]. Moreover, seven paper had more than 10 citations each [35, 55, 60, 62-64, 67]. In total, the citations number of all cited papers is 300, with an average of 6 citations per paper.

Table 10. show Number of google scholar citations

| Year of publication | Reference | Google Scholar Citation | Total number of Google scholar Citation |
|---|---|---|---|
| 2015 | [63] | 19 | 48 |
| | [71] | 2 | |
| | [64] | 17 | |
| | [38] | 21 | |
| | [67] | 17 | |
| | [68] | 8 | |
| 2016 | [57] | 3 | 116 |
| | [58] | 6 | |
| | [37] | 7 | |
| | [59] | 4 | |
| | [60] | 14 | |
| | [61] | 66 | |
| | [62] | 16 | |
| 2017 | [49] | 0 | 87 |
| | [50] | 2 | |
| | [66] | 0 | |
| | [73] | 3 | |
| | [51] | 2 | |
| | [52] | 2 | |
| | [35] | 14 | |
| | [53] | 6 | |
| | [8] | 6 | |
| | [24] | 1 | |
| | [70] | 3 | |
| | [54] | 4 | |
| | [55] | 10 | |
| | [56] | 3 | |
| | [25] | 23 | |
| 2018 | [26] | 1 | 13 |
| | [46] | 3 | |
| | [47] | 0 | |
| | [6] | 0 | |
| | [18] | 2 | |
| | [39] | 0 | |
| | [5] | 1 | |
| | [65] | 0 | |
| | [42] | 1 | |
| | [17] | 2 | |
| | [22] | 2 | |
| | [69] | 1 | |
| | [48] | 0 | |
| | [23] | 0 | |
| 2019 | [43] | 0 | 0 |
| | [7] | 0 | |
| | [40] | 0 | |
| | [20] | 0 | |
| | [21] | 0 | |
| | [41] | 0 | |
| | [44] | 0 | |
| | [45] | 0 | |
| | [36] | 0 | |

## 4. The Geographical Distribution of the Selected Studies

Figure 10 presents the nations that are most active in the ASA research area. We recognized 19 unique countries in ASA research from our primary studies. Saudi Arabia, with 18 published articles, is as of now the most dynamic nation out of the 19 countries in the research area, followed by Jordan with eight published articles. Egypt and the USA are the third most active countries with five published articles for each. Moreover, MOROCCO comes 5th with four published articles. Algeria and Canada are the 6th most active countries with three published articles for each. We highlighted five countries with two publications. The rest of the nation has less than two publications each.

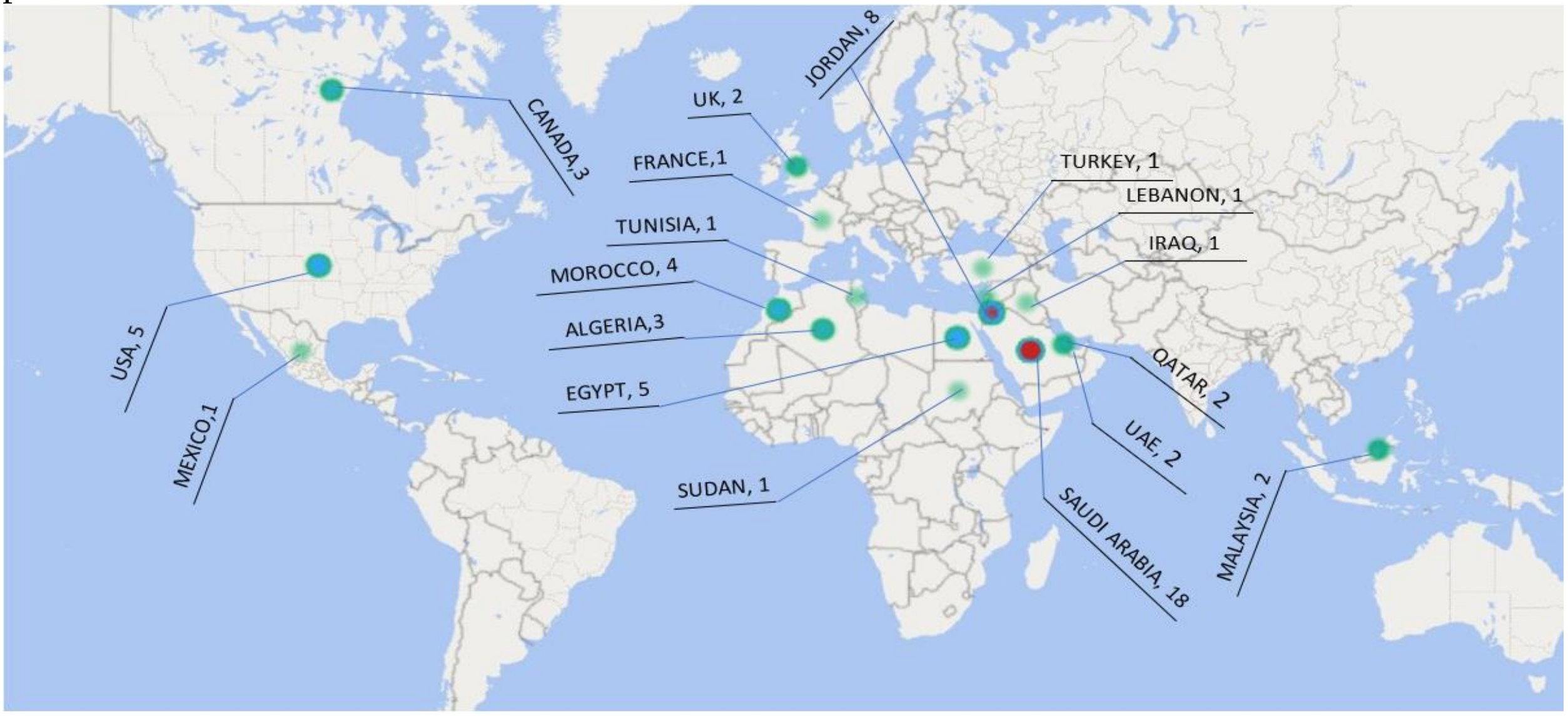

Fig 10. Map of most Active Countries in ASA studies.

## 5. Authorship Information

To comprehend the authors who had contributed the most as far as the number of publications from our 51 primaries papers, we collected the authors who have a high number of contributions. Table 11 demonstrates the rundown of the top10 most influential authors in the ASA research area. Al-Kabi, M. N. from Zarqa University, Jordan has shown up in 4 publications. This makes the authors the most influential in the ASA research area. Followed by Al-Twairesh, Nora from King Saud University, Riyadh, Saudi Arabia, and Mohammad, Al-Ayyoub from Jordan University Science Technology, Jordan has appeared in 3 publications. Followed by Al-Khalifa, H., Al-Salman, AbdulMalik, and Al-Dossari, H. Z., from King Saud University, Saudi Arabia with an appearance in 6 publications two studies for each. Followed by Wahsheh, H. A., from King Khaled University, Abha, Saudi Arabia has appeared in 2 publications. Followed by Alsmadi, I. M., from Boise State University, USA, has appeared in 2 studies. Followed by Bin Hathlian, N. F., from the Nairyah University of Hafar Al-Batin, Saudi Arabia has appeared in 2 publications. From our primary papers, we discovered 45 unique institutions from different active countries.

Table 11. Top 10 Most Influential Authors in ASA Research

| No | Author Name | Institute | Number of Paper | References |
|---|---|---|---|---|
| 1 | Al-Kabi, M. N. | Zarqa University, Jordan | 4 | [57, 60, 62, 67] |
| 2 | Al-Twairesh, Nora | King Saud University, Riyadh, Saudi Arabia | 3 | [5, 35, 39] |
| 3 | Mohammad, Al-Ayyoub | Jordan university science technology, Jordan | 3 | [60, 62, 63] |
| 4 | Jararweh, Yaser. | Jordan university science technology, Jordan | 2 | [62, 63] |
| 5 | Wahsheh, H. A. | King khaled university, Abha, Saudi Arabia | 2 | [57, 67] |
| 6 | Alsmadi, I. M. | Boise state university, USA | 2 | [57, 67] |
| 7 | Al-Khalifa, H., | King Saud university, Riyadh, Saudi Arabia | 2 | [35, 71] |
| 8 | Al-Salman, AbdulMalik., | King Saud university, Riyadh, Saudi Arabia | 2 | [35, 71] |
| 9 | Al-Dossari, H. Z., | King Saud university, Riyadh, Saudi Arabia | 2 | [7, 50] |
| 10 | Bin Hathlian, N. F., | Nairyah university of Hafar al Batin, Saudi Arabia | 2 | [24, 59] |

Table 12 shows the top 10 most active academic institutions in the field of study with respect to their number of publications. King Saud University, Riyadh, Saudi Arabia has the greatest number of publications with ten articles, followed by Jordan University of science technology with five publications, and Zarqa University, Jordan with five publications. However, the rest of the universities has two studies for each, which is three universities from Saudi Arabia (King Khaled, Nairyah, and Al-Imam Muhammad ibn Saud) and two universities from UAE (Ajman and Sharjah), and Qatar University and Princess Sumaya University for technology from Jordan.

Table 12. Top 10 most Active Institutions in ASA Research

| No | Organizations | Number of Paper |
|---|---|---|
| 1 | King Saud university, Riyadh, Saudi Arabia | 10 |
| 2 | Jordan University of science technology, Jordan | 5 |
| 3 | Zarqa University, Jordan | 4 |
| 4 | Ajman University, UAE | 2 |
| 5 | University Sharjah, UAE | 2 |
| 6 | Qatar University, Doha, Qatar | 2 |
| 7 | King khaled university, Abha, Saudi Arabia | 2 |
| 8 | Princess Sumaya University for technology, Jordan | 2 |
| 9 | Al imam Muhammad ibn Saud Islamic university, Saudi Arabia | 2 |
| 10 | Nairiyah university of Hafar al Batin, Saudi Arabia | 2 |

## H. The main findings, research challenges identified, and future work

This section was structure into two subsections. First, the principal findings of this study were highlighted in detail. Secondly, some of the identified research challenges from our PSS were highlighted and discussed with recommendations for future work direction.

1. *Main Findings:* The main aim of this study is to investigate the current study in the ASA research area. In doing so, 51 primary studies were selected based on our methodology to be our primary studies for analysis. The following are the main findings of our study.
    a. The most frequently used in Arabic sentiment analysis is the Machine learning technique, which is used in 76% of the selected studies. However, we found growing activities in both lexicon-based and hybrid techniques. Another observation is the increasing number of different studies, with 24% in total out of all the primary studies. However, we observed that the Machine learning technique is still consistently gaining attention, whereby from (2015 – 2019), 76% of the studies are based on ML technique. Probably we would see more work on lexicon-based and hybrid technique in years to come because they got a promised result in the last few years.
    b. The supervised learning approach is the most frequently used in Arabic sentiment analysis is the Machine learning technique, which is used in 70% of the selected papers. However, unsupervised approach and semi-supervised approach are seen growing activities ASA research area. Furthermore, 18% of the selected studies used the unsupervised approach, while the semi-supervised approach has 12%. Moreover, the observation is the increasing number of another approach used in ASA studies. However, we observed that unsupervised learning approach is consistently gaining attention, in the last few years because of new machine

learning algorithms not need training set such as Deep Learning. Probably we would see more work on unsupervised approach in years to come.

c. In ASA research area, we found out that 75% of our selected primary studies are used word level in a supervised approach. Another observation is the slowly increasing number of miscellaneous studies, with 25% in total out of all the primary studies. Sentence level and document level showed 17%, and 8% respectively.

d. On our research facet, 49% of the selected studies are evaluation papers, 47% are solution papers, and 4% are validation papers. Moreover, on contribution facet, Build Lexicon has 12 papers with 23%, followed by model and tools with 22%, the method with 12%, approach with 10%, framework four studies with 8%, and lastly, technique with 4%.

e. We observed that datasets are the most collected form Twitter, which is 60% data in ASA field of research, which is 23 papers of the selected studies. Hence, the Twitter platform is the most utilized by the researcher. In general, social media networks dataset are the most utilized which have some of the smallest used of dataset among all the dataset platforms utilized in our primary studies. However, recent researchers are shifting to more realistic and streaming datasets with big data sentiment analysis.

f. On our evaluation metrics, we observed that accuracy, recall, and precision expense score are the most utilized evaluation metrics by our selected studies. However, 9% of the evaluation metrics in primary selected studies were included as other evaluation metrics. Therefore, our analysis showed the top 10 evaluation metrics in the selected studies.

g. The top publication fora's as identified from our primary papers are Journal of Procedia Computer Science and International Journal of Advanced Computer Science with 13 and six papers respectively. Information Processing and Management, Expert Systems with Applications, ACM Transactions on Asian and Low-Resource Language Information Processing (TALLIP), Egyptian Informatics Journal, Journal of King Saud University – Computer and Information Sciences, International Journal of Advanced and Applied Sciences, International Journal on Semantic Web and Information Systems, and Pattern Recognition Letters have made an immense contribution with 8 publications form top 10 journal.

h. ASA research area has increased expanding consideration from the data science community since 2015, with an increasing number of publications with, on average, 45 publications from respectable journals and conferences every year. From our primary papers, we saw that about 59% of the research was published in journals, while 41% were distributed in conferences. With consistency in publication and increasing enthusiasm from the research community, we trust that the ASA area would presumably increase considerably more consideration the future.

i. Because of the high activity in the ASA research area, 6% of the chose papers to have more than 20 citation counts. As far as nations with high publications, we have recognized that 29% of our primary papers were distributed from Saudi Arabia, while 13% of the studies were from Jordan. However, countries like the United States with Egypt (8%), Morocco 6%, and 5% for Algeria with Canada. Moreover, Jordan is published respectable papers out of 19 countries. Therefore, other institutions in the rest of the countries should do more effort to move the ASA research area forward in the near future. Despite Jordan being the second most active country in ASA research, the most influential author (Al-Kabi) is from a Zarqa University, Jordan, with four publications, followed by Al-Twairesh, Nora from the King Saud University, Riyadh, Saudi Arabia and Mohammad, Al-Ayyoub from Jordan University Science Technology, Jordan with 3 publications for each. We have noticed that of all our most influential writers, 45% are not from the most active countries (Saudi Arabia).

2. *Research Challenges and Future Work:* After a thorough analysis of all our selected studies, we identified few research problems that the research community needs to address in ASA research domain and correspondingly, the recommendation for future work were given to guide researchers. We found out that data collected from the Twitter platform is the most utilized dataset in ASA research domain comparing with other social media platforms. However, the utilization of Twitter platform dataset is recently not considered sufficient anymore. This problem limits the generalization of the ASA results in the research domain. Therefore, for future work, researchers adopting more realistic and more extensive datasets is vital.

3. Moreover, the Arabic dialect dataset used for sentiment analysis, most of which was collected from Arab countries in Asia. Note that the Arab country in Africa represents 50% of the Arab countries. However, this dialect Arabic need more focus in future research. Considering the development of primary studies, we identified three main research type: evaluation, validation, and problem-solving. The research on the ASA seems to have high relevance on evaluation, which was evident in a large number of evaluation papers (49% of the primary studies) and a relatively decent number of problem-solving papers (47%). However, there was considerably less validation research (4%) in the literature. Therefore, this is a significant challenge in the ASA research area. Hence, for future work, more validation research is needed in the ASA research domain. Furthermore, some of the mentioned research challenges in the literature are also not sufficiently addressed despite much research effort. Research challenges such as Arabic dialect with lexicon-based because of lack of Arabic dialect dictionary[74], Building standard lexicon for Arabic dialect and modern standard Arabic [75] adopt a sentiment analysis methodology that works with different Arabic dialects[76].

   To address these concerning issues, new standard dataset/techniques/approaches/ frameworks etc. need to largely address them. Perhaps the application of more deep learning techniques might aid to address these vital challenges. The ASA, through the application of machine learning, stands out when looking at methods and algorithms. Moreover, traditional sentiment analysis methods and algorithms, like SVM, NB, DT, and K-means, are frequently applied, and researches tend to enhance the text representation by applying NLP and deep learning methods or using large external datasets. Therefore, for future work, researchers are adopting DL and NLP for MSA and incorporated into the sentiment analysis process. Considering the development of primary studies, we identified three main ASA levels: document level, sentence level, and word or aspect level. We expect an increase in the number of studies that have document level because it showed only four studies under a supervised approach, which is 8%. We also expect a raise of resources (linguistic resources and annotated corpora) for Arabic languages. These resources are essential for the development of ASA technologies. The greater availability of Arabic resources will provide more studies on these languages.

# 5. Conclusion

The Arabic language started takes place in sentiment analysis research since it has a significant impact on text meaning. Be that as it may, there is an absence of secondary studies that consolidate these researchers. This study an SMS directed to a review of Arabic concerned sentiment analysis literature. The scope of this study is broad for the first search (277 papers matched the search expression). Therefore, the mapping was chiefly performed dependent on abstracts and critical reading of papers.

By and by, we trust that our limitations do crucially affect the outcomes since our research has wide coverage. The principal contributions of our study are (1) it exhibits a quantitative analysis of the research ASA; (2) its pursued a well-characterized convention for SMS; (3) it discusses the area of ASA in regards to seven crucial aspects: the technique, machine learning approaches, level of sentiment analysis, Research facets, contribution facets, applications platforms, and demographic characteristics); and (4) the created mapping can contribute an overall outline of the ASA and can be of extraordinary assistance for conducting researchers in sentiment analysis and text mining. In this way, this effort can fill a gap in the literature as supposedly, to the best of our knowledge, this is the first general systematic mapping study of this extensive topic for ASA. Although a few kinds of research have been developed in the sentiment analysis field the processing of Arabic text remains an open research issue. The field needs secondary studies in areas that have a few numbers of primary studies, such as lexicon-based and hybrid approach.

An extra point in Arabic language concern. We found extensive differences in quantities of concentrates among modern standard Arabic and Arabic dialect since 56% of the recognized studies manage MSA. In this manner, there is an absence of studies managing dealing with texts written in Arabic dialect. When considering Arabic dialect text mining, we trust that this need can be filled up with the development of good knowledge bases and natural language processing methods specific to these languages. In addition, the sentiment analysis effect of dialects languages is additionally a motivating open research question in light of the variance of Arabic dialect. A correlation among feature aspects of various dialects and their effect on the results of sentiment analysis methods would be interesting.